\documentclass[pre,amsmath,amssymb,showpacs,groupaddress,twocolumn]{revtex4}
\usepackage{graphicx}
\usepackage{bm}
\begin{document}
\title{Self-Organized Criticality as Witten-type Topological Field Theory with Spontaneously Broken Becchi-Rouet-Stora-Tyutin Symmetry}
\author{Igor V. Ovchinnikov}
\email{iovchinnikov@ucla.edu, igor.vlad.ovchinnikov@gmail.com}
\affiliation{Department of Electrical Engineering, University of California at Los Angeles, Los Angeles, CA, 90095-1594}
\altaffiliation{Also at Laboratory for Nanophysics, Institute for Spectroscopy of Russian Academy of Science, Troisk, 142190, Russia.}
\begin{abstract}
Here, a scenario is proposed, according to which a generic self-organized critical (SOC) system can be looked upon as a Witten-type topological field theory (W-TFT) with spontaneously broken Becchi-Rouet-Stora-Tyutin (BRST) symmetry. One of the conditions for the SOC is the slow driving noise, which unambiguously suggests Stratonovich interpretation of the corresponding stochastic differential equation (SDE). This, in turn, necessitates the use of Parisi-Sourlas-Wu stochastic quantization procedure, which straightforwardly leads to a model with BRST-exact action, \emph{i.e.}, to a W-TFT. In the parameter space of the SDE, there must exist full-dimensional regions where the BRST-symmetry is spontaneously broken by instantons, which in the context of SOC are essentially avalanches. In these regions, the avalanche-type SOC dynamics is liberated from overwise a rightful dynamics-less W-TFT, and a Goldstone mode of Fadeev-Popov ghosts exists. Goldstinos represent modulii of instantons (avalanches) and being gapless are responsible for the critical avalanche distribution in the low-energy, long-wavelength limit. The above arguments are robust against moderate variations of the SDE's parameters and the criticality is "self-tuned". The proposition of this paper suggests that the machinery of W-TFTs may find its applications in many different areas of modern science studying various physical realizations of SOC. It also suggests that there may in principle exist a connection between some of SOC's and the concept of topological quantum computing.
\end{abstract}
\maketitle
\section{Introduction}
\label{Introduction}
Self-organized criticality (SOC) \cite{BWT} is rightfully considered by many to be a very fundamental phenomenon. SOC found its applications in geophysics, \cite{SOCGeology} neuro- \cite{SOC_Neurology, SOC_Neurology_exper} and evolutionary \cite{SOCBiologicalEvolution} biology, cosmology \cite{SOC_Cosmology} and astrophysics, \cite{SOC_Astro} collective human (traffic flow, \cite{TrafficFlow} stockmarkets \cite{StockMarket}) and animal \cite{SOC_Animal} behavior, cellular automation, \cite{CellularAutomata, CriticalCellularAutomata} and many other areas of modern scientific research. \cite{OtherVariousModels,BookJansen} Previous investigations have firmly established several distinct conditions for and properties of SOC, \cite{BookJansen} which we discuss first.

In Ref. \cite{BookJansen}, SOC has been given another, yet more definitive name of \emph{slowly driven, interaction dominated threshold} systems. Expanding on this definition, SOC is observed in highly nonlinear systems, which possess large number of metastable states and which are driven by slow external stochastic noise. The noise pumps the energy into the system, which thus must also be capable of damping the excess energy, \emph{i.e}, the system is dissipative. What has been just said can be reformulated into the two following conditions for the SOCs that are thus expected to appear in

(C1) stochastic nonlinear dissipative systems with large number of metastable states, when

(C2) the external noise is "slower" than the internal processes (scale separation principle).

SOCs possess a unique set of properties. The most important of them are the following three:

(P1) The time evolution is an infinite sequence of jumps between metastable states, \emph{i.e.}, of avalanches

(P2) with algebraic correlations (criticality),

(P3) which persist on moderate variations of the system's parameters (self-tuning).

It is not known yet if there exist a mathematical construction that starts with conditions (C1) and (C2) and reproduces properties (P1)-(P3). In this paper, it is proposed that such construction does exist and is known under the name of Witten-type topological field theories (W-TFT). A generic SOC state must be identified as a W-TFT with spontaneously broken Becchi-Rouet-Stora-Tyutin (BRST) symmetry.

The idea that a theory of SOCs must possess fermionic symmetries is not new. One of the known members of the SOC family is spin glasses. \cite{SpinGlasses} The glass phase of spin-glasses, which is always critical, was identified \cite{SpinGlassSusy} as the supersymmetry (susy) broken state of an N=2 susy theory, obtained from Parisi-Sourlas-Wu quantization of a Langevin stochastic differential equation (SDE). Moreover, the N=2 susy theory of self-organizing systems has also been proposed, \cite{SelfOrganizationSUSY} again based on the stochastic quantization of Langevin SDEs. At the same time, it is understood that a generic SOC system corresponds not to a Langevin SDE but rather to a general form SDE. Here, Parisi-Sourlas-Wu method is applied to general form SDEs, which is seemingly the only original part of the paper. For its most part, however, the paper is merely a collection of already established results, compiled into a physical picture, which we believe stands behind all SOC phenomena.

We begin in Sec. \ref{SOCSDE} with the discussion of a physical difference between Stratonovich and Ito interpretations of SDEs. We show that the scale-separation principle of SOC unambiguously suggests Stratonovich interpretation. In Sec. \ref{ParisiWuMain}, we demonstrate that the corresponding Parisi-Sourlas-Wu stochastic quantization leads to a (pseudo-Hermitian) model with a BRST-exact action, \emph{i.e.}, to a W-TFT as discussed in Sec. \ref{SDEasWTFT}. In Sec.\ref{Localization}, we discuss the localization principle and the need for a mechanism that breaks the BRST-symmetry and liberates SOC dynamics. In Sec. \ref{Instantons}, we show that this mechanism is instantons that in the context of SOC are essentially the avalanches. In Sec. \ref{ConditionsBRSTBreaking}, we speculate that BRST-symmetry breaking must occur in full-dimensional regions of the SDE's parameter space. In Sec.\ref{ScaleInvariance}, we switch to higher-dimensional SDEs and bring up the standard argument that the low-energy, long-wavelength part of the liberated SOC dynamics is the Goldstone Fadeev-Popov ghosts, which are gapless and thus are responsible for the critical distribution of avalanches (instantons). We argue that the "self-tuning" property is in the possibility to moderately vary the SDE parameters without hindering all the above reasonings. In Sec.\ref{Discussion}, we make a few remarks on the proposed scenario. We conclude in Sec.\ref{Conclusion}.

\section{SOC as SDE}
\label{SOCSDE}
\subsection{Physical meaning of SDE}
Condition (C1) in the Introduction is essentially the statement that the natural starting point for the studies of SOC's are nonlinear dissipative SDEs. Consider an SDE for $N$ stochastic variables:
\begin{eqnarray}
\partial_t\varphi^i(t) + A^{i}(t) = \xi^i(t),\label{GeneralSDE01}
\end{eqnarray}
where $i=1...N$, $A^i\equiv A^i(\varphi)$ is the vector field, which could be called the grift term, and $\xi^i$ is the stochastic noise. Metastable states correspond to the critical points of the drift term. Consequently, for SDEs representing SOC systems there is a large number of critical points $\#\left\{\alpha| A^{i}(\varphi_{\alpha})=0\right\}\gg1$. This is yet another way of saying that the system under consideration is highly nonlinear.

The drift term can also be given as
\begin{eqnarray}
\label{GeneralDriftTerm}
A^i = V^{'i} + {\tilde A}^i,
\end{eqnarray}
where the Langevin part can be defined via a potential, $V$, $V^{'i}\equiv\delta^{ij}V_{'j}, V_{'j}\equiv \partial V/\partial \varphi^j$ (the summation over the repeated indices is assumed throughout the paper), and the non-potential (magnetic, Hamilton) part, $\tilde A^i$, is such that ${\tilde A}^{i'j} \equiv \delta^{jk}\tilde A^i_{'k} = -\tilde A^{j'i}$. For the arguments of this section it suffice to consider the Euclidian target manifold. In the following sections, we will generalize the discussion to Riemannian target spaces.

That the system is dissipative suggests that $V$ is non-zero. The potential part of the drift term is responsible for the tendency of the open SOC system to dissipate its energy into a reservoir and minimize the potential $V$, while the noise stochastically pumps the energy from (yet another) reservoir. The dynamics represented by Eq.(\ref{GeneralSDE01}) has the physical meaning of a stochastic energy flow through a (highly non-linear) SOC system from an energy source to a drain.

The stochasticity comes into the system only from the source. This assumes that we have already out-integrated the drain's degrees of freedom. A drain reservoir with a memory can in fact introduce a temporal non-locality into the SDE. In the lowest order approximation this must have the following form: $A^i \to \int_{-\infty}^{t} M^i_j(t-t')A^j(\varphi(t'))dt' $, where $M$ could be called a memory kernel. The kernel has a certain characteristic time, $\lambda_d, M^i_j(t-t') \to 0$ as $|t-t'|/\lambda_d\to\infty$. In our interpretation of the scale separation principle, $\lambda_d$ is (one of) the shortest time-scale(s) in the problem. Under this condition, one can always work on a scale much larger than $\lambda_d$, on which the drain reservoir is memory-less and is rightfully represented as the Langevin part of Eq.(\ref{GeneralDriftTerm}).

Note also, that the "source" reservoir's noise, $\xi$, may not necessarily be temporarily local (white). In particular, our argument toward Stratanonvich interpretation of the SDE in the next subsection does not rely on the assumption of the "whiteness" of the source noise.

For many (if not all) systems identified as SOC, the non-potential part of the drift term is also nonvanishing. Therefore, we are interested in cases when both $V\ne0$ and $\tilde A^i\ne 0$. Throughout the paper we refer to this situation as to a general form SDE as opposed for example to the case of Langevin stochastic differential equation, $\tilde A^i=0$, corresponding to Witten model, \cite{WittenMorse} which is the (0+1)-dimensional beginning of all the W-TFTs.

\subsection{Stratonovich vs. Ito interpretations of SDE}
\label{StratonovichvsIto}
SDEs can be treated on field-theoretic grounds. At this, the quantization procedure depends on interpretation of stochastic noise. There are two major choices: Stratonovich and Ito interpretations. They are related respectively to the stochastic quantization procedures of Parisi-Sourlas-Wu \cite{ParisiWu, StochasticAsTopological, ParisiWuPlusOneDimension} and of Martin-Sigia-Rose. \cite{MartinSigiaRose} The choice depends on physical conditions. In this section we show that the scale separation principle (condition (C2) in the Introduction) unambiguously suggests the Stratonovich picture.

Consider the discrete version of Eq.(\ref{GeneralSDE01}): \cite{ItoStratonovich}
\begin{eqnarray}
(\varphi_{t+\epsilon}^i-\varphi^i_t)/\epsilon + ((1-\zeta)A^{i}_{t+\epsilon}+\zeta A^{i}_t) = \xi^i_{t+\epsilon},\label{LangevinDiscrete}
\end{eqnarray}
where time now takes on discrete values separated by $\epsilon$, the noise on the rhs should be thought of as that acting during time interval $[t,t+\epsilon]$, and $\zeta$ is the parameter that essentially represents the speed of noise in comparison with internal processes, \emph{e.g.}, to equilibration. Indeed, when the system is slower than the noise, it does not have enough time to adjust its variables to the quickly changing noise. Therefore, the infinitesimal change in $\varphi_t$ within each time step $[t,t+\epsilon]$ must be determined by the earliest value of $A^{i}$, \emph{i.e.}, by $A^i_t$. Consequently, fast noises correspond to the Ito case with $\zeta=1$.

In the opposite case of fast system, the Stratonovich choice of $\zeta=1/2$ is natural. The infinitesimal change in $\varphi_t$ is determined by $A^{i}$ "averaged" over the time-interval $[t,t+\epsilon]$, or rather by its value in the middle of $[t,t+\epsilon]$, \emph{i.e.}, by $(A^i_{t+\epsilon}+A^i_t)/2$. \cite{footnote2} This essentially means that the system has enough time to adjust its variables to the noise, before the slow noise changes considerably.

The only statistical partition function in the model is that of the noise:
\begin{eqnarray}
\mathcal{Z} = \int [\text{d}\xi] P(\xi),\label{PartitionFunctionNoise}
\end{eqnarray}
where $P$ is the noise distribution function. If we impose periodic boundary conditions on $\varphi$, however, the numbers of $\varphi_t$'s and $\xi_t$'s will be the same. In this case the partition function can be rewritten as a path-integral over $\varphi$'s instead of $\xi$'s by the appropriate variable transformation:
\begin{eqnarray}
\mathcal{Z} \stackrel{\xi\to\varphi}{\longrightarrow} \int [\text{d}\varphi] J(\zeta) P(\xi(\varphi)),\label{PartitionFunction}
\end{eqnarray}
where $\xi(\varphi)$ is the lhs of Eq.(\ref{LangevinDiscrete}) and the Jacobian of the variable transformation, $J(\zeta) = |\partial \xi_t/\partial \varphi_{t'}|$, is
\begin{eqnarray*}
J(\zeta) = \left|\begin{array}{cccccc}
\hat D_0    &0         & 0             & \dots &0             &\hat N_\text{T}   \\
\hat N_0    &\hat D_\epsilon& 0             & \dots &0             &0     \\
0      &\hat N_\epsilon& \hat D_{2\epsilon} & \dots &0             &0     \\
\vdots &\vdots    & \vdots        & \ddots&\vdots        &\vdots\\
0      &0         &0              &\dots  &\hat D_{\text{T}-\epsilon}& 0    \\
0      &0         &0              &\dots  &\hat N_{\text{T}-\epsilon}& \hat D_\text{T}
\end{array}\right|,
\end{eqnarray*}
with $\hat D_t = \epsilon^{-1}\delta^{i}_{j}+(1-\zeta)A^{i}_{'j}$ and $\hat N_t = -\epsilon^{-1}\delta^{i}_{j}+\zeta A^{i}_{'j}$.

Note that noise is typically assumed Gaussian and physics contained in Eq. (\ref{PartitionFunctionNoise}) is trivial. On the other hand, Eq.(\ref{PartitionFunction}) has appeared from Eq. (\ref{PartitionFunctionNoise}) by the formal redefinition of the variables. Therefore, if the physics of Eq.(\ref{PartitionFunction}) is not trivial this can only be blamed on the non-trivial topology (\emph{e.g.}, not one-to-one) of the highly nonlinear (and nonlocal) map, $\varphi:\xi^i(t)\to\varphi^i(t)$. \cite{Nicolai} This is the first indication on the topological nature of the Parisi-Sourlas-Wu stochastic quantization procedure, the core of which as compared to Martin-Sigia-Rose procedure is in "not forgetting" the Jacobian of the variable transformation. The physical justification for this appreciation of the Jacobian follows.

On taking the continuous limit one obtains (up to a $\epsilon$-dependent constant)
\begin{eqnarray}
J(\zeta)\stackrel{\epsilon\to0}\longrightarrow e^{(1-\zeta) \int_{t=0}^\text{T} A^{i}_{'i}}-e^{-\zeta \int_{t=0}^\text{T} A^{i}_{'i}}.\label{ItoTrivialization}
\end{eqnarray}

For Ito interpretation, when $\zeta=1$, the first term in rhs of Eq.(\ref{ItoTrivialization}) is a constant. Furthermore, one can assume that due to the tendency of the system to minimize the potential, it spends most of its time in the region(s) where $A^{i}_{'i}=V^{'i}\text{}_{i}>0$. Therefore, $\int_{t=0}^\text{T} A^{i}_{'i}\stackrel{\text{T}\to \infty} {\longrightarrow}+\infty$, the second term is negligible, and one sets $J(1)\stackrel{\text{T}\to \infty} {\longrightarrow}1$. Consequently, if the noise is fast one can neglect the Jacobian.

The scale separation principle, however, suggests that for SOCs, the Stratonovich interpretation of stochasticity is more appropriate. Recall that in the sand-pile model, before one adds yet another grain to a random site, he has to time-propagate the system until it finds its new metastable state. Among the physical examples are earthquakes which are instantaneous when compared to the slow process of tension build-up in the earth crust due to plate tectonics, while the relatively long periods of (quasi-)equilibrium in the punctuation theory of biological evolution are followed by sudden reconfigurations. Therefore, for the studies of SOC we must use Stratonovich interpretation of SDEs and keep the Jacobian with $\zeta=1/2$.

One may expect, however, that for realistic situations $\zeta$ is not exactly $1/2$. To see how $\zeta>1/2$ can change our story it is convenient to rewrite the Jacobian as:
\begin{eqnarray}
J(\zeta) = e^{-(\zeta-1/2)\int_{t=0}^\text{T} A^{i}_{'i}} J_{P},\label{GeneralJ}
\end{eqnarray}
where $J_P = 2\sinh ((1/2)\int_{t=0}^\text{T} A^{i}_{'i})$ can (and will be) represented as a path integral over additional fermions (ghosts) with periodic boundary conditions (see below). $\zeta>1/2$ explicitly breaks BRST-symmetry of the model. \cite{SoftSusyBreaking} The ghosts become energetically more costly due to the additional term $(\zeta-1/2) A^{i}_{'i}$ in the Lagrangian. This may lead to considerable changes in the low-energy dynamics even for small $\zeta-1/2\ll1$. Therefore, the Stratonovich interpretation of SDE is of cause an approximation. It describes a hypothetical SOC system, for example a numerical model, for which the noise can be assumed infinitely slow. Having said that, we always assume $\zeta=1/2$ in the rest of the paper.

So far we did not specify the noise weighting function, $P$. From previous discussion, a reasonable approximation for $P$ of a slow noise is a temporarily non-local Gaussian with considerably large noise-noise correlation time, $\lambda_s$. At the same time, in the following sections we will use "white" noise as the driving force ($\lambda_s\to0$). This may seem a contradiction with the scale-separation principle. There is no contradiction though. One can straightforwardly generalize the developments in the following sections to temporarily nonlocal noises, which, however, will turn out later to be an unnecessary complication. The point is that we can always rescale time $t\to\Lambda t$ so that $\lambda_s\to\lambda_s/\Lambda\to 0$. \cite{footnote3} From this perspective, the scale separation principle is solely in the appropriate choice of the stochastic quantization procedure.

\section{Parisi-Sourlas-Wu quantization of SDE}
\label{ParisiWuMain}
\subsection{Path-integral approach}
\label{BRSTExactness}
Let us now proceed with the case of Gaussian white noise and with the Parisi-Sourlas-Wu quantization \cite{footnote4} of Eq.(\ref{GeneralSDE01}), applicability of which is based on the appreciation of the importance of the Jacobian corresponding to the Stratonovich interpretation of SDE in the previous section. The partition function is:
\begin{eqnarray}
\mathcal{Z}&=& \int [\text{d}\varphi] J_{P} e^{-\int_{t=0}^\text{T} G_{ij}K^i K^j/2}, \label{InitialPartitionFunction}
\end{eqnarray}
where $K^i = \partial_t\varphi^i + A^{i}$ and $G^{ij}=(G_{ij})^{-1}$ is the noise-noise correlator,
\begin{eqnarray}
\langle\xi^i(t)\xi^j(t')\rangle = G^{ij} \delta(t-t'),\label{Noise01}
\end{eqnarray}
which for now is assumed to be independent of $\varphi$ (see below for the covariant generalization). One can always think of $G_{ij}$ as of the metric of the target manifold as long as $A^i$ is arbitrary. In other words, there is no Riemannian structure on the target manifold yet, and (\ref{Noise01}) could as well play the role of the metric. It will be seen later that it actually does play this role.

The Jacobian can be represented as the path integral over the fermionic Fadeev-Popov ghosts according to $\det M = \int [\text{d}\chi][\text{d}\bar \chi]\exp(\bar\chi_i M^{i}_{j}\chi^j)$. Furthermore, one can employ the Legandre multiplier, $B_i$, which is the dynamical conjugate to $\varphi$. The partition function now is:
\begin{eqnarray}
\mathcal{Z}= \int [\text{d}\Phi]e^{-S},\label{EuclidianPartitionFunction}
\end{eqnarray}
where $\Phi$ represents all the fields, and the action $S=\int_{t=0}^\text{T} L$ is defined by the Lagrangian
\begin{eqnarray}
L = iB_i(K^{i} - i G^{ij}B_j/2) - \bar\chi_i(\partial_t \chi^i + A^{i}_{'j}\chi^j). \label{EuclidianLagrangian}
\end{eqnarray}
The periodic boundary conditions are imposed on all the fields. In case of ghosts, these conditions are needed if the fermion determinant is to represent the Jacobian.

The model enjoys the global nilpotent fermionic BRST-symmetry ($\mathcal{Q}$-symmetry) \cite{StochasticQexact} given by the infinitesimal operator
\begin{eqnarray}
\label{QOperator}
\mathcal{Q}  = \sum\nolimits_i\int_{\text{t}} \chi^i(t) \delta/\delta\varphi^i(t) + iB_i(t) \delta/\delta\bar\chi_i(t),\label{BRST}
\end{eqnarray}
so that ${\mathcal Q}S \equiv \left\{{\mathcal Q},S\right\} = 0$. Importantly, the action is also $\mathcal Q$-exact:
\begin{subequations}
\label{QExactTheory01Dimensions}
\begin{eqnarray}
S = \left\{{\mathcal Q},\Psi\right \}, \label{Q-exactness}
\end{eqnarray}
where
\begin{eqnarray}
\Psi = \int_{t=0}^\text{T} \bar\chi_i (K^{i} - i G^{ij}B_j/2),\label{GaugeFermion}
\end{eqnarray}
\end{subequations}
is known as the gauge fermion. $\mathcal Q$-exact action is a unique feature of W-TFTs. It looks like the whole action is nothing else but the BRST gauge-fixing term. In Ref.[\onlinecite{book1}], this was identified as the "quantizing zero" situation.

From the field-theoretic point of view, a more appealing derivation of (\ref{QExactTheory01Dimensions})  is based on the philosophy that the SDE itself can be looked upon as the gauge choice for the two "independent" stochastic variables - $\varphi$ and $\xi$:
\begin{eqnarray}
\mathcal{Z}= \int [\text{d}\varphi][\text{d}\xi][\text{d}\chi][\text{d}\bar\chi] e^{-\int_\text{t=0}^\text{T} G_{ij}\xi^i\xi^j/2 + ...},
\end{eqnarray}
where dots denote the gauge-fixing $\mathcal{Q}$-exact term, $\{\mathcal{Q}, \int_\text{t=0}^\text{T} \bar \chi_i(K^i-\xi^i)\}$ (here $\mathcal{Q}$ has the form (\ref{BRST}) with $iB_i\to G_{ij}\xi^j$). By noticing that $G_{ij}\xi^i\xi^j/2=\{\mathcal{Q},\bar\chi_i \xi^i/2\}$ is also $\mathcal{Q}$-exact and by formal redefinition, $\xi^i\to iG^{ij}B_j$, one recovers (\ref{QExactTheory01Dimensions}).

Moreover, the above Parisi-Sourlas-Wu quantization can be generalized to the cases when the noise-noise correlator (the metric of the target manifold) is dependent of $\varphi$, $G^{ij}\to G^{ij}(\varphi)$. The Batalin-Vilkovisky procedure must be used to come up with the same $\mathcal Q$-exact action (\ref{QExactTheory01Dimensions}). \cite{StochasticQexact,book1} At this, however, the $\mathcal Q$ operator (\ref{QOperator}) will acquire a more intricate form, which accounts for the curvature of the target manifold. The detailed explanation of the corresponding quantization procedure and the appropriate form of the $\mathcal Q$-operator can be found in Appendix A.2 of Ref.[\onlinecite{book1}], where the only adjustment to our case needed to be done is $G^{ij}V_{'j} \to A^i$.

After the covariant generalization, Eq. (\ref{QExactTheory01Dimensions}) represents a very general (0+1) SDE, which is quantized stochastically by the Parisi-Sourlas-Wu method in accordance with the Stratonovich interpretation of the slow noise.

\subsection{Schr\"odinger picture}
\label{Schroedeinger}
Let us turn now to the Schr\"odinger picture. The statistical (Euclidian) partition function (\ref{EuclidianPartitionFunction}) can be given through the Hamiltonian function as :
\begin{eqnarray}
\mathcal{Z} = \int [\text{d}\Phi] e^{ \int_{t=0}^T (i \pi_{\varphi^i} \partial_t\varphi^i + i \pi_{\chi^i} \partial_t\chi^i - H)},\label{generalHamiltonian}
\end{eqnarray}
where $\pi$'s are the canonical momenta, which on passing to the Schr\"odinger picture $\pi_{\varphi^i}\to -i\partial_{\varphi^i}$ and $\pi_{\chi^i} \to - i\partial_{\chi^i}$. From Eqs.(\ref{generalHamiltonian}) and (\ref{EuclidianLagrangian}) we identify $\pi_{\varphi^i}= -B_i$ and $\pi_{\chi^i} =i\bar\chi_i$, so that in flat coordinates:
\begin{eqnarray}
H &=& -\triangle/2 - [A^i,\partial_{\varphi^i}]_+/2 + A^i_{'j} [\partial_{\chi^i},\chi^j]_-/2, \label{HamiltonianFlat}
\end{eqnarray}
where $\triangle=\partial_{\varphi^i} G^{ij}\partial_{\varphi^j}$ is the Laplacian. The choice of the operator ordering in the second term of Eq.(\ref{HamiltonianFlat}) is standard. The operator ordering in the last term has direct connection to the Stratonovich interpretation of the noise of the SDE. Have we been considering "faster" noises, we would have to use $\zeta A^i_{'j}\partial_{\chi^i}\chi^j -(1-\zeta) A^i_{'j}\chi^j \partial_{\chi^i}$ instead of the last term in Eq.(\ref{HamiltonianFlat}), that is we would have to add $(\zeta-1/2) A^i_{'i}$ to the Hamiltonian. (c.f., Eq.(\ref{GeneralJ}))

N\"oether charge associated with the $\mathcal{Q}$-symmetry is:
\begin{eqnarray}
Q = -iB_i\chi^i \to \chi^i\partial_{\varphi^i},\label{Exterior}
\end{eqnarray}
which is nothing else but the exterior derivative on the target manifold. This is so far the second explicit revealing of the topological nature of the model. As it should, the charge is nilpotent, $Q^2=0$, and commutative with $H$.

The Hamiltonian can also be given as:
\begin{eqnarray}
H = \{ Q,\bar Q\}/2, \label{AntiCommutator}
\end{eqnarray}
where
\begin{eqnarray}
\bar Q = \bar \chi_i(iG^{ij}B_j + 2A^i) \to \partial_{\chi^i}(-G^{ij}\partial_{\varphi^j} - 2 A^i). \label{barQOperator}
\end{eqnarray}
In non-flat coordinates, the Hamiltonian is again an anticommutator (\ref{AntiCommutator}), while explicitly:
\begin{eqnarray}
H = -\triangle /2 - \mathcal{L}_{A^i},\label{Hamiltonian}
\end{eqnarray}
where the Laplacian is given by the Weitzenb\"ock formula that includes four-fermion coupling through the Riemmann curvature tensor, $\mathcal{L}_{A^i}$ is the Lie derivative along $A^i$. N\"oether charge is the same, while $\bar Q = \star Q\star - 2\mathfrak{i}(A^i)$ with $\star$ and $\mathfrak{i}(A^i)$ denoting the Hedge operation and interior multiplication by $A^i$.

Hamiltonian (\ref{Hamiltonian}) has a very clear physical meaning. The first term is the quantum mechanical "smearing" (dispersion) of wave-functions, which are p-forms from the (complexified) exterior algebra of the target manifold. The intensity of the dispersion is determined quantitatively by the "magnitude" of noise-noise correlator, $||G^{ij}||\sim\text{temp}$, which thus has the meaning of noise temperature. This is the essence of the stochastic quantization, the stochasticity takes the form of the quantum mechanical fluctuations. In the classical, low-temperature limit, \cite{ClassMechanics} one is left with a non-dispersive classical flow of p-forms along the drift term, $A^i$.

Hamiltonian (\ref{Hamiltonian}) is not Hermitian with respect to the conventional metric in the Hilbert space: \cite{book1,book2}
\begin{eqnarray}
\langle\alpha|\beta\rangle = \int (\star \alpha^*)\wedge\beta,\label{ConventionalMetric}
\end{eqnarray}
where $\wedge$ is wedge product of p-forms and the integration is over the target manifold. All entries in the Hamiltonian, however, are real. Therefore, the Hamiltonian can be looked upon as an infinite-dimensional real matrix. The spectrum of such Hamiltonian consists of real energies or pairs of complex-conjugate energies. This means that the Hamiltonian is pseudo-Hermitian \cite{Mostafazadeh} and there must exist such hermitian, invertible $\eta$ that
\begin{eqnarray}
H^\dagger = \eta H\eta^{-1}.
\end{eqnarray}
Model (\ref{QExactTheory01Dimensions}) is a pseudo-Hermitian quantum mechanics \cite{Bender, Mostafazadeh} with $\eta$ being the metric of the Hilbert space
\begin{eqnarray}
\langle\langle\phi|\psi\rangle\rangle \equiv\langle\phi| \eta \psi\rangle, \langle\langle\phi| \equiv \langle\phi|\eta, |\psi\rangle\rangle \equiv |\psi\rangle,
\end{eqnarray}
preserved by the Schr\"odinger evolution:
\begin{eqnarray}
i\partial_t\langle\langle\phi| \psi\rangle\rangle = \langle\langle\phi| H - \eta^{-1}H^\dagger\eta |\psi\rangle\rangle =0.
\end{eqnarray}
For Witten model, $\eta_W=e^{2V}$. In fact, Witten model is a quasi-Hermitian quantum mechanics. The eigenvalues of its Hamitonian are all real, since $\eta_W^{1/2}H\eta_W^{-1/2}$ (such transformation brings Eq.(20) to its conventional form appearing in the literature) is Hermitian. For a general form SDE, $\eta$ is complicated and highly non-local, and the energies are complex.

The anticommutator form of the Hamiltonian (\ref{AntiCommutator}) is that of the N=2 pseudo-supersymmetric (p-susy) quantum mechanics, \cite{PseudoSupersymmetry} with the following operator algebra:
\begin{subequations}
\label{SUSYconditions}
\begin{eqnarray}
&H_0 = \{Q_0,Q_0^\sharp\}/2, Q_0^2 = [H_0,Q_0] =0,\label{SUSYconditions1}\\
& (Q_0^\sharp)^2 = [H_0,Q_0^\sharp] =0.\label{SUSYconditions2}
\end{eqnarray}
\end{subequations}
and $Q_0^\sharp=\eta^{-1} Q_0^\dagger \eta$ is the pseudo-Hermitian conjugate to $Q_0$. In order for the p-susy to be unbroken, the operator algebra (\ref{SUSYconditions}) must be complemented with the ground state(s) such that
\begin{eqnarray}
Q_0|0\rangle\rangle = Q_0^\sharp|0\rangle\rangle =0.\label{SUSYGround}
\end{eqnarray}

N=2 p-susy is twice larger than the $\mathcal Q$-symmetry. It is a combination of $\mathcal Q$-symmetry and yet another fermionic pseudo-anti-$\mathcal{Q}$ symmetry related to $\mathcal Q$ by the pseudo-time-reversal conjugation ($\eta\mathcal{T}$-conjugation).

In general case, however, model (\ref{QExactTheory01Dimensions}) does not possess the pseudo-anti-$\mathcal Q$ symmetry. Indeed, $\bar Q$, which is supposed to play the role of $Q_0^\sharp$, is neither nilpotent nor does it commute with $H$. With $\bar Q$ for $Q^\sharp_0$, the operator algebra Eqs.(\ref{SUSYconditions2}) fails. For example, in flat coordinates:
\begin{eqnarray}
&(\bar Q)^2 = 2 F^i_k G^{kj}\partial_{\chi^i}\partial_{\chi^j}\ne 0,\label{barQ2}
\end{eqnarray}
where $F^i_{j} = (A^i_{'j}-A^j_{'i})/2$ is the "field tensor" of the non-potential (magnetic) part of the drift term.

If for $F^i_j\ne 0$ the N=2 p-susy is present, it means that
\begin{eqnarray}
(\bar Q)^2,[H,\bar Q]|\text{phys}\rangle\rangle =0,\label{WierdCondition}
\end{eqnarray}
are satisfied for all the (physical) states of the model. For these situations, one can find such  $\eta$ that $\bar Q = \eta^{-1} Q^\dagger \eta$. These situation may appear, for example, if $A^i$ is a Killing or a symplectic vector field. \cite{HamiltonSUSY,HamiltonAndLangevin} There is no reason, however, to believe that this is true in general. Below we will argue that in the SDE's parameter space there must exist region(s) with explicitly broken N=2 p-susy (together with the spontaneously broken $\mathcal Q$-symmetry).

In the rest of the paper we use the path-integral representation of the pseudo-Hermitian quantum mechanics. This is advantageous for the reason that finding the exact form of the metric, $\eta$, is a complicated task and one should avoid $\eta$ whenever possible. Fortunately, the metric is automatically incorporated into the path-integrals and one must not worry about its explicit form. \cite{NonHermitianPathIntegral}

\section{SDE as W-TFT}
\subsection{Conditions for W-TFT}
\label{SDEasWTFT}
To identify a theory as a W-TFT, one needs: \cite{book1}

(\emph{i}) a $\mathbb{Z}_2$-graded nilpotent fermionic $\mathcal Q$,

(\emph{ii}) a $\mathcal Q$-exact action,

(\emph{iii}) a $\mathcal Q$-invariant path-integral measure, and

(\emph{iv}) a $\mathcal Q$-closed ground state(s), \cite{Fujikawa} which ensures that the $\mathcal Q$-symmetry is not spontaneously broken.

These conditions suffice to establish a unique set of properties of W-TFTs. \cite{book1} In particular, one can introduce a "metric" for time: $\text{d}t\to e(t)\text{d}t$, and in case of higher-dimensional theories the metric for spatial dimensions, $\hat g$ (our case is lacking Lorentz invariance so that $e$ and $\hat g$ should not be combined into a space-time metric). In result, the gauge fermion acquires explicit dependence on $e$ and $\hat g$: $\Psi\to\Psi(e,\hat g)$. The topological nature of W-TFTs is seen through the $(e,\hat g)$-independence of (now we consider T$\to\infty$ limit corresponding to the field-theoretic interpretation of the path-integral):
\begin{eqnarray}
\langle\langle0| \mathcal{T} A B ... |0\rangle\rangle = \int [\text{d}\Phi] e^{-\{\mathcal{Q},\Psi\}} A B ... ,\label{TopInvariants}
\end{eqnarray}
where $A,B...$ are $\mathcal {Q}$-closed, \emph{i.e.}, $\{\mathcal {Q},A\}, \{\mathcal {Q},B\}...=0$, and $\mathcal{T}$ denotes chronological ordering. Indeed, the functional variation with respect to, \emph{e.g.}, $e(t)$:
\begin{eqnarray}
\frac{\delta}{\delta e(t)} \langle\langle0| \mathcal{T} A B ... |0\rangle\rangle = \langle\langle0| \mathcal{T} \{\mathcal{Q},\gamma_e(t)\} A B ...|0\rangle\rangle=0, \label{QClosedCorrelators}
\end{eqnarray}
where $\gamma_e(t) = -\delta\Psi/\delta e(t)$. The last equality can be proven by partial integration ($\mathcal Q$ is a differentiation operator) and holds only if the ground state(s) obey (in the Sch\"odinger picture)
\begin{eqnarray}
Q|0\rangle\rangle=Q^\sharp|0\rangle\rangle=0,
\end{eqnarray}
with $Q^\sharp=\eta^{-1}Q^\dagger\eta$. We see that this condition (and/or condition (\emph{iv}) above) is actually that for the unbroken N=2 p-susy (\ref{SUSYGround}).

We can not speak with certainty that the breakdown of N=2 p-susy always tailors the spontaneous breaking of $\mathcal Q$-symmetry. There may exist situations when $Q|0\rangle\rangle=0$, while $Q^\sharp|0\rangle\rangle\ne0$. Such situations are at least exotic. We omit these exotic situations in this paper and think of the two symmetry breakdowns as of equivalent.

In fact, the proof of that the $\mathcal Q$-symmetry is unbroken is a subtle and complicated part of W-TFTs. \cite{book2,WittenSUSYBreaking} Therefore, we find ourselves now in a fortunate position that the $\mathcal Q$-symmetry breaking is actually what we are looking for. To demonstrate the breakdown of the $\mathcal{Q}$-symmetry of (\ref{QExactTheory01Dimensions}) it suffice to establish, for example, the T-dependence of the partition function. Indeed, the partition function is one of the topological invariants (\ref{TopInvariants}) with $AB...$ being a unity operator that is obviously $\mathcal Q$-closed. This suggests that $\mathcal Z$ of a true W-TFT is time-deformation invariant and thus T-independent. On the other hand, model (\ref{QExactTheory01Dimensions}) definitely satisfies conditions (\emph{i})-(\emph{iii}) above so that the spontaneous $\mathcal Q$-symmetry breaking is the only possible source for the T-dependence of $\mathcal{Z}$ of otherwise rightful W-TFT with T-independent $\mathcal{Z}$.

\subsection{Localization principle}
\label{Localization}
Let us turn now to the (lowest order) one-loop study of partition function. Consider the case of low temperatures ($||G^{ij}||\ll1$) so that $A^i$ is very "pronounced". The major contribution into the partition function comes from the lowest energy states that cluster around critical points of $A$ (the metastable states of the SOC). This can be seen by out-integrating the Legandre multiplier, $B$:
\begin{eqnarray}
\label{SimplifiedAction}
\mathcal{Z} = \int[\text{d}\Phi] e^{-\int_\text{t=0}^\text{T} K^iG_{ij}K^j/2+\text{fermions}}.
\end{eqnarray}
The contribution is that from the fluctuations around the stationary paths, $\varphi^i_\text{cl}(t)=\varphi^i_\alpha$, so that:
\begin{eqnarray}
\mathcal{Z}^\text{(1-loop)} = \sum\nolimits_\alpha \mathcal{Z}_\alpha\label{Ztotal}.
\end{eqnarray}
${\mathcal Z}_\alpha$ can be calculated in the locally Euclidian coordinates, $\delta\varphi^i$, in which the Gaussian part of the action is:
\begin{eqnarray}
S^{(2)} = \int_{t=0}^\text{T}\left(\delta \varphi^i (\hat {\mathcal D}_-\hat {\mathcal D}_+)^{ij}\delta\varphi^j/2 + \bar\chi^i \hat {\mathcal D}_+^{ij}\chi^j\right),\label{GaussianAction}
\end{eqnarray}
where $\hat {\mathcal D}_+ = \partial_t + \hat a, \hat {\mathcal D}_- = -\partial_t + \hat a^T$ with $a^{ij}$ being $A^{i}_{'j}(\varphi_\alpha)$ in the $\delta\varphi$-coordinates. The Gaussian integration leads to the bosonic and fermionic determinants:
\begin{eqnarray}
\mathcal{Z}_{\alpha}=|\hat{\mathcal D}_+|/|\hat{\mathcal D}_- \hat{\mathcal D}_+|^{1/2}.
\end{eqnarray}

Introducing temporal Furrier components, \emph{e.g.}, $\delta\varphi^i = \sum_{n=-\infty}^{\infty}\delta \varphi^i(n)e^{i \omega_n t}, \omega_n = 2\pi n/\text{T}$, one gets
\begin{eqnarray*}
\mathcal{Z}_{\alpha}&=&\Xi_{\alpha}\times |\hat a|/|\hat a\hat a^T|^{1/2},\nonumber\\
\Xi_\alpha&=&\prod\limits_{n>1}\frac{|\omega_n^2+\hat a^2|}
{(|\omega_n^2 + \hat a^2||\omega_n^2 + (\hat a^T)^2|)^{1/2}}.
\label{DeltaZ2}
\end{eqnarray*}
To simplify further this expression one notices that since $\hat a$ is real, its eigenvalues, $a_i, i=1...N$, are either real or come in complex conjugate pairs, and that the set of eigenvalues of $\hat a^T\equiv a^\dagger$ is $a_i^*, i=1...N$. Therefore, the sets of eigenvalues and consequently the determinants of $\omega_n^2+\hat a^2$ and $\omega_n^2 + (\hat a^T)^2$ are the same. Hence, $\Xi_\alpha=1$ and:
\begin{eqnarray*}
\mathcal{Z}_\alpha = (-1)^{\Delta_\alpha},
\end{eqnarray*}
where $\Delta_\alpha$ is the number of real negative $e_n$'s. $(-1)^{\Delta_\alpha}\equiv \text{sign}|A^i_j(\varphi_\alpha)|=\text{ind}_\alpha$ is known as the index of the critical point, and
\begin{eqnarray}
\mathcal{Z}^\text{(1-loop)} = \sum\nolimits_{\alpha} \text{ind}_\alpha,\label{OneLoop}
\end{eqnarray}
is the Euler characteristic of the target manifold according to the Poincar\'e-Hoft theorem. Hence, on the one-loop level, the fluctuations leave the $\mathcal Q$-symmetry intact as is seen from the T-independence of Eq.(\ref{OneLoop}).

The reason for bringing the reader's attention to the one-loop calculation that does not break the $\mathcal Q$-symmetry is twofold. Firstly, Eq.(\ref{OneLoop}) is yet another explicit demonstration of the topological nature of the model. Secondary, it reveals the tendency of ghosts to compensate for bosonic fluctuations. Up to one-loop, the ghosts completely cancel fluctuations in the bosonic fields which means that so far the model has no dynamics. This compensation is known in the literature as the localization principle. As the name suggests, the path integral is "localized" to classical solutions of the SDE. Stationary solutions at critical points is one class of classical solutions. The other class is instantons (see next subsection).

From the physical point of view, the meaning of the localization principle is that the slow noise does not provide the system with "frequencies". In result, the dissipative system does not fluctuate.

For Witten model, the one-loop approximation is known to be exact from the perturbative point of view. For a general form SDE, however, higher-order fluctuational corrections may in principle break the $\mathcal Q$-symmetry (anomalously in that sense that perturbatively provided corrections possess lower symmetry than the original action). \cite{footnoteUnlikely} However, even if the fluctuations leave the symmetry intact, it can still be broken by instantons. \cite{InstantonBrerakingOfSusy, book1, book2, WittenSUSYBreaking} In fact, instantons is the primarily source of the $\mathcal Q$-symmetry breakdown as we discuss in the next section. In particular, instantons break the $\mathcal Q$-symmetry even of Witten model itself, when the Euler characteristic of the target manifold is zero. \cite{book1} It is the instanton-induced $\mathcal Q$-symmetry breakdown which we identify as an SOC and which we are interested in.

\subsection{Meaning of instantons and their role in $\mathcal Q$-symmetry breaking}
\label{Instantons}
\begin{figure}[t] \includegraphics[width=8.6cm,height=3.5cm]{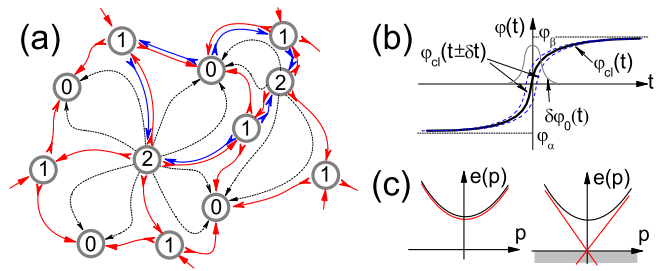}
\caption{\label{Figure1} (Color online) {\bf (a)} Schematics representing an SDE's drift vector field, $A^i$, and its critical points (circles) with their indices given explicitly. Thick double-arrowed curves represent instantons (classical solutions of SDE connecting two critical points). They are complemented by the axillary thin vector-lines which connect critical points with indices differing by 2. Thick "reversed" double-arrowed curves (given for only one closed path) represent antiinstantons, which as is explained in the text have different matrix elements from those of the corresponding instantons due to the non-potential (magnetic) part of $A^i$. This imbalances instantons and antiinstantons on time-reversed loops, \emph{e.g.}, the inner and outer loops shown. Under certain conditions, this imbalance must lead to instanton-induced spontaneous BRST-symmetry breaking. The argument is similar to that of Anderson localization of a particle on a random lattice. In the presence of a magnetic field, the constructive interference of time-reversed hopping loops, which is responsible for the localization, fails. {\bf (b)} In the semi-classical approximation, each instanton is a Gaussian path-integral around classical solution of the SDE (thick center curve) connecting two critical points ($\varphi_\alpha$ and $\varphi_\beta$) at $t=\pm\infty$. This solution has the so-called modulus - its center. The variation in the modulus, \emph{e.g.}, $\varphi_\text{cl}(t)\to \varphi_\text{cl}(t\pm\delta t)$, does not change the action of the instanton. This leads to the existence of a zero fluctuational mode, $\delta\varphi_0(t) = \partial_t\varphi_\text{cl}(t)$. The corresponding zero ghost mode decouples from the action and the path-integral is zero unless this zero mode is matched. The matrix element of $\mathcal Q$-operator is non-zero because it also has an additional ghost operator matching the zero ghost mode. By similar arguments, instantons can only connect critical points with indices differing by $1$ (in case of antiinstantons, $-1$) as is seen from figure (a). {\bf (c)} The qualitative difference between excitation spectra of higher-dimensional theories with unbroken (left) and spontaneously broken (right) $\mathcal Q$-symmetry. The double curve on the left represents dispersions of supersymmetric partners. The cone on the right represents goldstinos. The negative energy states (shaded) are occupied and form a (gap-less) Dirac sea. The Dirac-sea ground state represents a quantum liquid of solitons (textures, patterns) as explained in the text.}
\end{figure}

The SOC dynamics can be described as follows. The system spends most of its time in its metastable states, while the time evolution is an infinite sequence of sudden jumps (avalanches) between the metastable states. In our terms, the metastable states are nothing else but the perturbative ground states (PGS), $|\alpha\rangle\rangle$, around critical points of $A^i$. The "jumps" between different PGS's are the processes of the quantum mechanical tunneling, which are called instantons. Therefore, in the application to the SOC systems, \emph{instantons are nothing else but the mathematical term for the avalanches.} \cite{AvalanchesExplanaition}

Instantons in a sense are the opposite of fluctuations. The processes of quantum mechanical tunneling are exponentially weak as compared with fluctuations. That instantons are of low-energy means that they do not happen often. Fluctuations are localized around critical points unlike instantons which connect them. Hence, instatons are of ultimate nonlinear character. In linear situations they simply do not exist. This is yet another sense in which instantons are qualitatively different from fluctuations. Furthermore, as it will be clear later from the discussion of higher-dimensional models, avalanches are the only low-energy, long-wavelength dynamics in the system. This is especially appealing for the physical picture of SOCs.

The above picture of SOC makes sense only when the critical points possess distinct PGS's. In other words, when the PGS's do not overlap much. The overlap can be provided by a sufficient "smearing" of the PGS's due to large kinetic energy. In the model under consideration, the kinetic energy in (\ref{Hamiltonian}) is determined by the noise temperature (the magnitude of the metric $G^{ij}$). Therefore, for an SOC to occur (at least in its recognizable form) \emph{the noise must not only be slow but also weak}.

The low-energy instanton dynamics can be defined by the projection onto the reduced Hilbert space spanned by $|\alpha\rangle\rangle$'s: \cite{book1,book2}
\begin{eqnarray*}
\mathbb{H}_{\alpha\beta} &=& \{\mathbb{Q},\bar{\mathbb{Q}}\}_{\alpha\beta}/2,\\
\mathbb{Q}_{\alpha\beta} &=& \langle\langle \alpha| Q |\beta\rangle\rangle, \bar{\mathbb{Q}}_{\alpha\beta} = \langle\langle \alpha| \bar Q |\beta\rangle\rangle,
\end{eqnarray*}
where $\mathbb{Q}$ and $\bar{\mathbb{Q}}$ are matrix elements of instantons and antiinstantons.
In the path-integral language,
\begin{subequations}
\label{braket}
\begin{eqnarray}
|\alpha\rangle\rangle = \int [\text{d}\Phi] e^{-\int_0^{t=+\infty} L},
\end{eqnarray}
where the functional integration is over the paths starting at $t=0$ at the argument of $|\alpha\rangle\rangle$ and ending at $t=+\infty$ at $\varphi^i_\alpha$. The (pseudo-) bra is accordingly:
\begin{eqnarray}
\langle\langle\alpha| = \int [\text{d}\Phi] e^{-\int_{t=-\infty}^0 L},
\end{eqnarray}
\end{subequations}
where the integration is over the paths starting at $t=-\infty$ at $\varphi^i_{\alpha}$ and ending at $t=0$ at the argument of $\langle\langle\alpha|$.

From Eqs.(\ref{braket}), the instanton and antiinstanton matrix elements can be given as:
\begin{subequations}
\begin{eqnarray}
\label{InstantonsPathIntegral}
\mathbb{Q}_{\alpha\beta} &=& \int [\text{d}\Phi] Q(0) e^{-\int_{t=-\infty}^{+\infty} L},\\
\bar{\mathbb{Q}}_{\alpha\beta} &=& \int [\text{d}\Phi] \bar Q(0) e^{-\int_{t=-\infty}^{+\infty} L},
\end{eqnarray}
\end{subequations}
where the integration is along the paths connecting the two critical points, $\varphi^i(-\infty)=\varphi^i_\alpha$ and $\varphi^i(+\infty) = \varphi^i_\beta$.

In the one-loop approximation, the matrix elements of instantons can be found by Gaussian integration around a classical solution connecting the two critical points, $\partial_t\varphi^i_\text{cl}(t)+A^i(\varphi_\text{cl}(t))=0, \varphi^i_\text{cl}(-\infty)=\varphi^i_\alpha,\varphi^i_\text{cl}(+\infty)=\varphi^i_\beta$. As in Sec.(\ref{Localization}), the Gaussian integration can be done in the locally Euclidian coordinates for $\varphi$, so that:
\begin{eqnarray}
\mathbb{Q}_{\alpha\beta} &=& \int [\text{d}\Phi] Q(0) e^{-S^{(2)}},\label{InstantonGaussian1}
\end{eqnarray}
with $S^{(2)}$ from Eq.(\ref{GaussianAction}). This time, however, operators $\mathcal{D}_\pm$ have explicit time-dependence, \emph{e.g.}, $\hat{\mathcal{D}}_+ = \partial_t + \hat a (\varphi^i_\text{cl}(t))$. If the classical solution connects critical points with indices differing by one, $\text{ind}\alpha = \text{ind}\beta+1$, operator $\hat{\mathcal{D}}_+$ has one zero mode. Indeed, due to the time-translation invariance, the classical solution can be shifted in time $\varphi^i_\text{cl}(t)\to \varphi^i_\text{cl}(t+\delta t)$ and still remain the classical solution. Differentiation of SDE with respect to $\delta t$ results in $\hat{\mathcal{D}}_+\partial_t\varphi^i_\text{cl}(t) = 0$. This mode corresponds to the infinitesimal shift of the center of the instanton, which is called instanton modulus. Furthermore, there is the corresponding ghost mode which decouples from the fermionic action. Therefore, the path-integral contains an integration over an unmatched Grassmann number. Such path integral must be zero unless the matrix element is that of an operator of ghost number 1. $Q$ is such an operator and its matrix element is non-zero.

Consider now the classical solution which connects critical points with indices differing by more than one, \emph{e.g.}, $\text{ind}\alpha = \text{ind}\beta+2$, represented by thin lines in Fig.(\ref{Figure1})a. Such solution can be though of as an instanton followed by yet another instanton. Each instanton comes with its own modulus and with the corresponding zero ghost mode. The path-integral has two unmatched Grassmann numbers. The matrix element of the ghost number 1 operator $Q$ is zero. Hence,
\begin{eqnarray}
\mathbb{Q}_{\alpha\beta} \ne 0, \text{ind}\alpha = \text{ind}\beta +1.\label{InstantonValue}
\end{eqnarray}

Similar reasonings apply to antiinstantons, which correspond to the backward propagation in time. Gaussian integration must be performed around a classical solution of the "time-reversed" differential equation, $\partial_t\bar\varphi^i_\text{cl}(t)-A^i(\bar\varphi_\text{cl}(t))=0, \bar\varphi^i_\text{cl}(-\infty)=\varphi^i_\alpha,\bar\varphi^i_\text{cl}(+\infty)=\varphi^i_\beta$. The antiinstanton analogue of (\ref{InstantonGaussian1}) is
\begin{eqnarray}
\bar{\mathbb{Q}}_{\alpha\beta} &=& e^{- 2\Delta V_{\beta\alpha} - 2 \int \tilde A^i G_{ij} d\bar\varphi^j_\text{cl}}\int [\text{d}\Phi] \bar Q(0) e^{- S^{(2)}},\label{InstantonGaussian2}
\end{eqnarray}
where $\Delta V_{\beta\alpha}=V(\varphi_\beta)-V(\varphi_\alpha)>0$ and $\tilde A$ is the non-potential part of $A$ (see Eq.(\ref{GeneralDriftTerm})). Now, it is $\hat{\mathcal{D}}_-$ which has a zero mode and the corresponding $\bar\chi$ must be matched, which $\bar Q (0)$ actually does so that:
\begin{eqnarray}
\bar{\mathbb{Q}}_{\alpha\beta} \ne 0, \text{ind}\alpha + 1= \text{ind}\beta.
\end{eqnarray}
As compared to Eq.(\ref{InstantonGaussian1}), Eq.(\ref{InstantonGaussian2}) has two additional elements. The first one is $\Delta V_{\beta\alpha}$, which has the meaning of the potential difference at two critical points. It provides an antiinstanton with the exponentially weak tunneling factor. The factor can be absorbed into the pseudo-Hermitian metric of the reduced Hilbert space by the appropriate rescaling of PGS's: $|\alpha\rangle\rangle\to e^{V(\varphi_\alpha)}|\alpha\rangle\rangle,\langle\langle\alpha|\to \langle\langle\alpha|e^{-V(\varphi_\alpha)}$. In Witten model, this will result in that $\mathbb{Q}_{\alpha\beta}=\bar{\mathbb{Q}}_{\beta\alpha}$. The other part, $\int \tilde A^i G_{ij} d\bar\varphi^j_\text{cl}$, has the meaning of the (imaginary) phase factor acquired by a particle moving in a vector potential, $\tilde A^i$, of a magnetic field. Due to this part, the matrix elements of instantons are different from those of the corresponding antiinstantons even if we try to rescale $|\alpha\rangle\rangle$'s. This imbalances instantons and antiinstantons. The hopping (instanton-antiinstanton) evolution along $\tilde A^i$ becomes preferable.

Importantly, the integrant in Eq.(\ref{InstantonGaussian1}) is $\mathcal Q$-exact: $Q(0)=-iB_i(0)\chi^i(0) = -\{{\mathcal Q}, \bar\chi_i(0)\chi^i(0)\}$. A non-zero average of a $\mathcal Q$-exact operator, (\ref{InstantonValue}), is a direct indication on the $\mathcal Q$-symmetry breakdown. Therefore, the $\mathcal{Q}$-symmetry is prone to be broken by instantons. \cite{InstantonBrerakingOfSusy} Thus, instantons (antiinstantons) is the primarily candidate for the $\mathcal Q$-symmetry breaking. In fact, this effect is so pronounced that even balanced instantons and antiinstantons of Witten model break $\mathcal Q$-symmetry if the Euler characteristic of the target manifold is zero. \cite{book1}

The identification of $\tilde A^i$ as a vector potential of a magnetic field brings about the analogy with the problem of Anderson localization (see Fig.\ref{Figure1}). Originally, \cite{Anderson} Anderson considered a quantum particle which lives on a random lattice (the "lattice" of the PGS in our case). The particle hops between the lattice sites (instantons and antiinstantons in our case). Due to the constructive interference between time-reversed paths, the probability to stay at the same site is always greater than that of traveling and if the conditions are right this leads to the localization (preservation of $\mathcal Q$-symmetry in our case). Once, however, one introduces the magnetic field into the system, the time-revered paths get imbalanced and the constructive interference argument fails thus leading to the delocalization ($\mathcal Q$-symmetry breakdown in our case).

The conclusion of this subsection is as follows. If in the low-temperature limit the $\mathcal Q$-symmetry is broken, it is most likely due to the instanton-antiinstanton imbalance induced by the non-potential part of the drift term. The low-energy part of the liberated dynamics is of instanton (avalanche) type and from the mathematical point of view is described by the low-energy ghost modes, which in turn represent modulii of instantons (avalanches).

From the quantum-mechanical treatment, however, it is not clear why the distribution of avalanches (instantons) must be a power-law. To see this one must turn to higher-dimensional theories as we do in Sec.\ref{ScaleInvariance}. Before that, however, a few words on the conditions for the $\mathcal Q$-symmetry breakdown are in order.

\subsection{BRST-symmetry breaking}
\label{ConditionsBRSTBreaking}
An important question is when the spontaneous breakdown of $\mathcal Q$-symmetry occurs. We do not have a satisfactory answer to this question. We believe, however, that the answer may have a lot to do with the interesting observation of Ref. \cite{ExplicitPSUSYBreaking}. There, it was suggested and demonstrated with a few examples that in $\mathcal{PT}$-symmetric pseudo-Hermitian quantum mechanics the explicit breakdown of N=2 p-susy is always accompanied by a spontaneous breakdown of $\mathcal{PT}$-symmetry. The breakdown(s) occurs when the parameters of a model reach some critical values and the pairs of complex-conjugate energies appear. $\mathcal{PT}$-symmetric models \cite{Bender} is a subclass of pseudo-Hermitian $\eta\mathcal T$-symmetric quantum mechanics. \cite{Mostafazadeh} Therefore, it is reasonable to expect that the proposition of Ref.\cite{ExplicitPSUSYBreaking} can be generalized to other $\eta{\mathcal T}$-symmetric models such as the one under consideration.

That the $\eta\mathcal T$-symmetry breaking is needed in our case can be seen again from the analogy with the Anderson localization problem. In order to delocalize an Anderson particle one needs to break the $\mathcal T$-symmetry (to imbalance $\mathcal T$-reversed paths) by the introduction of a magnetic field, whereas in our case one needs to break the $\eta\mathcal T$-symmetry by the introduction of the magnetic (non-potential) part of the drift term.

In our case, there is also a seeming indication on the necessity of the appearance of complex energies for the existence of the liberated dynamics. The point is that the Parisi-Sourlas-Wu stochastic quantization leads to a "statistical" (Euclidian) partition function. The operator of the quantum mechanical time-evolution has the form of the propagation in imaginary time:
\begin{eqnarray}
\hat U(t) = \sum\nolimits_n e^{-t E_n}|n\rangle\rangle\langle\langle n|,
\end{eqnarray}
where $n$ numerates the levels and $E_n$ are their energies. Imaginary time usually has the meaning of temperature. On the other hand, we know that the time in the evolution operator is the original time of the SDE and not a temperature of any kind (temperature in our case is the metric of the target manifold). Therefore, the time-evolution does not make a conventional quantum-mechanical sense. Presumably, this can be attributed to the absence of the propagating modes and/or to the unbroken $\mathcal Q$-symmetry.

When a pseudo-Hermitian Hamiltonian, however, possesses complex energies, $E_n\to E_n + i \tilde E_n$, the evolution operator becomes:
\begin{eqnarray}
\hat U(t) = \sum\nolimits_n e^{-i t \tilde E_n + ...}|n\rangle\rangle\langle\langle n|,
\end{eqnarray}
which is now more of the taste of a conventional quantum mechanical evolution. One can hypothesize that this is the result of the liberated dynamics and/or of the spontaneous breakdown of the $\mathcal Q$-symmetry.

The emerging picture is as follows. If we take a Witten model with unbroken $\mathcal Q$-symmetry and start changing the SDE's parameters, we will eventually reach some critical values at which $\mathcal Q$- and $\eta\mathcal{T}$-symmetries will spontaneously breakdown. This will be signified by the appearance of complex energies of the Hamiltonian.

Disregard of whether the speculations in this subsection are correct or not, it is natural to expect that \emph{SOC dynamics is liberated from otherwise a rightful W-TFT in some low-temperature full-dimensional regions of the SDE's parameter space}. At least for N=2 susy models of spin-glasses this statement seems to be correct, \cite{SpinGlassSusy} and there is no reason for the situation to be dramatically different in the more general case of N=2 p-susy.

\section{Higher-dimensions and Goldstone criticality}
\label{ScaleInvariance}
The fact that the $\mathcal Q$-symmetry is spontaneously broken suffice to establish the criticality of the avalanche dynamics. All what is said for (0+1)-theories can be generalized to higher dimensions. \cite{footnote5} The generalization can be viewed as a limit of an infinite-dimensional target manifold. Literally, the index $i$ is split into the spatial coordinates, $\bm x$, and the coordinates of the target manifold: $\varphi^i\to\varphi^i(\bm x)$.

The higher-dimensional counterpart of Eq.(\ref{GeneralSDE01}) is:
\begin{eqnarray}
\partial_t\varphi^i(\bm xt)+A^i(\varphi) = \xi^i(\bm xt),\label{SDE}
\end{eqnarray}
where $A^i(\varphi)$ is some functional of $\varphi$'s and the Gaussian noise correlates on the metric of the target manifold:
\begin{eqnarray}
\langle\xi^i(\bm xt)\xi^i(\bm x't')\rangle = g^{ij}(\varphi)\delta(t-t')\delta^d(\bm x-\bm x').
\label{Noise}
\end{eqnarray}
After the Parisi-Sourlas-Wu quantization, the action is $\mathcal Q$-exact and defined by the gauge fermion (\ref{GaugeFermion}) with the additional integration over the spatial coordinates, $\bm x$.

In situations when the $\mathcal Q$-symmetry is spontaneously broken there must exist a local operator $\rho(\bm x t)$ such that:
\begin{eqnarray}
\langle \langle \left\{\mathcal{Q},\rho(\bm xt) \right\}\rangle\rangle \ne 0. \label{NonQExactness}
\end{eqnarray}
By the standard argument, there is a gapless Goldstone ghost mode. \cite{WittenSUSYBreaking,SalamStrathdee} Consider the following average:
\begin{eqnarray}
\langle \langle \rho({\bm x} t) \rangle\rangle = \int [\text{d}\Phi] \rho(\bm x t) e^{-S}.
\end{eqnarray}
Under the path-integral we can make the space-time dependent infinitesimal transformation of the fields $\delta_\epsilon \Phi(\bm x' t') = \epsilon (\bm x' t') \left\{{\mathcal Q},\Phi(\bm x' t')\right\}$. The average is not to be changed since the transformation is merely the change in the integration variables: $\delta_\epsilon \langle \langle \rho({\bm x} t) \rangle \rangle = 0$ or
\begin{eqnarray}
\epsilon(\bm x t)\langle \langle \{\mathcal{Q},\rho({\bm x} t) \}\rangle\rangle = \langle \langle \rho({\bm x} t)\delta_\epsilon S\rangle\rangle.\label{variation}
\end{eqnarray}
By the $\mathcal{Q}$-exactness of the action:
\begin{eqnarray}
\delta_\epsilon S = \int_{\bm x' t'} \left((\partial_\mu \epsilon) J^\mu\right) ({\bm x}' t'),\label{Noether}
\end{eqnarray}
where index $\mu$ combines space-time coordinates and $J^\mu$ is the N\"oether current associated with $\mathcal{Q}$. $\epsilon$ can be chosen constant within some arbitrarily large space-time volume, $\Omega$, such that $(\bm x t)\in\Omega$, and $\epsilon =0$ outside the volume. From Eqs.(\ref{Noether}), (\ref{variation}), and (\ref{NonQExactness}) we get the integral form of the failed Ward-Takahashi identity:
\begin{eqnarray}
\oint_{\partial\Omega(\bm x' t')}\left\langle \langle \rho(\bm x t)J^\mu(\bm x' t') \right\rangle \rangle n({\bm x}' t')_\mu = \text{const} \ne0,\label{WardIdentity}
\end{eqnarray}
where the integration is over the boundary of $\Omega$ and $n_\mu$ is the unit vector normal to the boundary. As long as $\Omega$ can be chosen arbitrarily large, the above equation shows that $\langle\langle\rho(\bm x t) J^\mu(\bm x' t') \rangle\rangle$ falls off algebraically as a function of $\bm x'-\bm x$ and $t-t'$. This happened due to the existence of the gapless Goldstone ghosts (goldstinos), which are "struggling" to restore the broken symmetry of the vacuum.

The spectrum of excitations must look like the one given in Fig.\ref{Figure1}c. Goldstinos must form a gapless Dirac sea with negative energy states being occupied. Furthermore, the would-be bosonic superpartners of goldstinos (upper curve of Fig.\ref{Figure1}c) must also be liberated from the fermionic symmetry and in many cases of interest they must be gapped. Thus, goldstinos (representing avalanches, see below) is the only dynamics the system has in the low-energy, long-wavelength limit.

In fact, $\rho(\bm xt)$ in Eq.(\ref{NonQExactness}) is the spatial density of ghosts and due to the existence of the Dirac sea, $\langle\langle \rho(\bm x t) \rangle\rangle\ne0$. In Ref.\cite{SelfOrganizationSUSY}, $\langle\langle \rho(\bm x t) \rangle\rangle$ was identified with the entropy density. In this manner, the spontaneous $\mathcal Q$-symmetry breaking and the appearance of the corresponding Dirac sea is related to the concept of "entropy production".

As in case of supersymmetric quantum mechanics, \cite{book2} the ghosts of PGS's (which in higher-dimensional cases could also be called perturbative vacua) represent unstable directions of $A^i$ in the functional space of $\varphi(\bm x)$. To find the ghosts of a particular perturbative vacuum, one must diagonalize the linear operator $\left(\delta A^i/\delta \varphi^j({\bm x}')\right)$ at the corresponding static field configuration, $\varphi^i_\alpha(\bm x)$, such that $A^i(\varphi_\alpha)=0$. Those modes which have negative real parts of their eigenvalues are occupied by the ghosts.

In cases of interest, the drift term functional $A^i(\varphi)$ does not explicitly depend on the spatial position $\bm x$. Therefore, the fact that a perturbative vacuum has many unstable directions (ghosts) most certainly implies that $\varphi_\alpha(\bm x)$ is spatially inhomogeneous. Such vacua could be looked upon as textures, patterns, \emph{etc}. They can also be called configurations of solitons such as domain walls, vortices, \emph{etc}. The actual "Dirac sea" vacuum is a quantum mechanical superposition of perturbative vacua corresponding to various solitonic configurations. In other words, {\emph{the ground state(s) of an SOC is a quantum liquid of solitons (textures, patterns)}.

As compared to the case of quantum mechanics discussed previously, higher dimensional theories must have an important new element. Critical points of the functional $A^i$ are not isolated. \cite{footnote6} Solitons could be thought of as instantons in space (not time). Therefore, they also have modulii at least due to the symmetries of the space. For example, if the model is invariant with respect to translations in space, field configurations obtained from a given $\varphi^i_\alpha(\bm x), A^i(\varphi_\alpha)=0$ by all the spatial translations, $\varphi^i_{\alpha\bm X}(\bm x)=\varphi^i_\alpha(\bm x+\bm X)$, are also critical field configurations for $A^i$, $A^i(\varphi_{\alpha\bm X})=0$, and thus are also candidates for a perturbative vacua. In this case, Bott-Morse theory applies, \cite{book2} which states that the perturbative vacua are from the cohomology of the modulii space of solitonic configurations. For the soliton modulii of spatial translations, this corresponds to the zero-momentum vacuum.

As we discussed before, the low-energy, long-wavelength part of the liberated modes, which as we saw are the goldstinos, must represent modulii of instantons (avalanches) connecting different perturbative vacua. Therefore, the algebraic correlator of the gapless ghosts assumes critical distribution of avalanches. \emph{We believe this is the essence of the criticality of SOCs}.

From discussion in Sec.\ref{ConditionsBRSTBreaking} we know that in the parameter-space of the SDE, the model has $\mathcal Q$-symmetry spontaneously broken in regions of the same dimensionality as the parameter space itself. Therefore, for any SOC we have a freedom to moderately vary the SDE parameters leaving the system at the Goldstone criticality. \emph{We believe this is the essence of the self-tuning property of SOCs}.

\section{Discussion}
\label{Discussion}
In this section we would like to make a few remarks which seem interesting:

$\bullet$ Previous studies of SOC used Ito interpretation of SDEs. This approach led to the conclusion that SOCs are members of the family of non-equilibrium phase transitions such as directed percolation \cite{SCOtoDIRPERC} captured by the Reggeon field theory. \cite{DIRPERCtoRFT}

Such approach certainly fails to explain the self-tuning property of SOCs. The point is that on the phase-diagram, \emph{i.e.}, the SDE's parameter space, a conventional critical state separates phases of different qualities and thus occupies manifolds of lower dimensionality than the phase-space itself. In other words, there is always at least one direction in the phase space, which leads the system off its criticality. The picture proposed in this paper seems to resolve this issue by explaining the criticality of SOC by the Goldstone theorem. Note, that the Goldstone explanation of the "self-tuning" property of spin-glasses is known for both the N=2 susy approach \cite{SpinGlassSusy,SpinGlasses} and replica trick approach. \cite{GoldstonemodeInSpinGlass}

$\bullet$ Viewing SOCs through the prism of conventional critical states has yet another flaw - it relies significantly on the renormalization group methods, which are essentially perturbative. On the other hand, perturbative methods can not be straightforwardly applied to avalanches (instantons), which are inherently of ultimate non-perturbative nature. The perturbation theory, however, can be meaningfully applied to some collective variables. \cite{DoussalWiese,DSFisher,footnote7} In our case, such collective variables are the instanton modulii or rather the corresponding ghosts. The low-energy effective theory for the ghosts would provide instantons with a sort of dual fermionic description, which admits perturbative treatment. This may be an advantage of the W-TFT picture of SOCs.

$\bullet$ Soon after the introduction of the concept of SOC, it was proposed that SOCs may occur only in the so-called conservative models. \cite{ConservativevsNon,BookJansen} Whether the conservative-nonconservative classification is accurate is still under debate. \cite{SOCNonConservative,NonCriticalityOfNonconserv} The proposed W-TFT picture of SOCs did not rely on a specific form of the SDE. Hence, it may turn out that the SOC family is bigger that it is believed now. Some order-out-of-chaos-type systems \cite{OrderOutOfChaos} (\emph{e.g.}, pattern formations) may as well belong to the SOC family.

$\bullet$ The origin of the self-tuning property of the SOC rests on the necessity to use the topological completion of the stochastic quantization (Parisi-Sourlas-Wu quantization) due to the slow noise. This can be paraphrased in an informal yet seemingly accurate form: the criticality of SOC is "topologically" protected.

$\bullet$ An important ingredient of all the W-TFTs is a set of $\mathcal Q$-closed operators, from which one can construct topological invariants (\ref{TopInvariants}). Typically, these operators are cycles of different dimensionality in the base manifold and their explicit form depends on the field content of a theory and/or on topology of the target space.

The fact that the $\mathcal Q$-symmetry is broken may lead to the conclusion that these operators are not topological invariants anymore and thus are useless. This is not quite so. The point is that due to the localization principle, perturbative vacua are still $\mathcal Q$-closed (at least up to one-loop). Therefore, one can use these operators for the topological classification of the perturbative vacua. Having constructed such a classification, one may introduce a refined SOC time-evolution of the system as a sequence of jumps, at which the topology of the quantum state of the system suddenly changes. In general, however, not every physical avalanche changes the topology of the state.

$\bullet$ To our opinion, one of the most interesting directions of future investigations could as well be the search for a possibly existing connection between SOC's and the concept of fault-tolerant (topological) quantum computing. \cite{Kitaev} If exists, this connection would be an exciting news not only for physicists but also for neuroscientists.

$\bullet$ We find it very unorthodox, though adequate to think of earthquakes as of fermions.

\section{Conclusion}
\label{Conclusion}
In conclusion, let us sketch again the discussed scenario, which we believe is a likely candidate for a theory of a generic SOC behavior, and explicitly reveal the connections between the conditions for and properties of SOCs outlined in the Introduction. The condition of the slow external driving (C2) leads to the necessity of using the Stratonovich interpretation of noise in the stochastic differential equation representing an SOC (C1). This necessitates the topological "completion" of the stochastic quantization procedure (Parisi-Sourlas-Wu procedure) and leads to a model with BRST-exact action - to a Witten-type topological field theory. In the SDE's parameter space, there are full-dimensional regions, in which the BRST-symmetry is spontaneously broken and the SOC dynamics is liberated. The low-energy, long-wavelength part of the liberated SOC dynamics represents avalanches (instantons) (P1). The liberated SOC dynamics can also be viewed as the Goldstone ghosts, which have no gap and thus are responsible for the critical avalanche distribution (P2). The BRST-symmetry of the model and its breakdown, which are the essence of the criticality, can not be lifted by a moderate variation of the parameters of the model. This is the essence of the self-tuning property (P3).

To the best of our knowledge, so far Witten-type topological field theories have only been of "internal" mathematical use as a tool for the studies of topologies of lower-dimensional manifolds. The proposal of this paper suggests that Witten-type topological field theories may find their applications in many other areas of science that study various realizations of self-organized criticality, \emph{e.g.}, in geophysics, astrophysics, neuroscience, evolutionary biology, \emph{etc}.

\acknowledgements
We would like to thank Hsien-Hang Shieh, Per Kraus, Eric D'Hoker, Akhil Shah, Robert N. Schwartz, and Kang L. Wang for discussions and critical reading of the manuscript. The work was supported by Defense Advanced Research Projects Agency, Defense Sciences Office, Program: Physical Intelligence, under Contract HR0011-01-1-0008.


\begin{references}
\bibitem{Nicolai} In the high-energy physics nomenclature, such a variables' transformation is known as Nicolai map.
\bibitem{SoftSusyBreaking} The explicit breaking of supersymmetry is also known in high-energy physics as soft supersymmetry breaking.
\bibitem{footnote3} At this, we also have to rescale the drift term and the noise-noise correlator (see below) $A^i\to\Lambda A^i$ and $G^{ij}\to\Lambda G^{ij}$.
\bibitem{footnote4} Note, that in some cases, the Parisi-Sourlas-Wu quantization is used to relate a theory at equilibrium to a N=2 supersymmetric theory in plus one "fictitious" time dimension. \cite{ParisiWuPlusOneDimension} In our case, there is no extra dimension, and $t$ is the ordinary time of the SDE.
\bibitem{footnote6} In fact, the same complication may have appeared in case of quantum mechanics if we did not assume that critical points of $A^i$ are not isolated, but instead form some higher dimensional submanifolds of the target space.
\bibitem{footnote7} For example, in Ref.[\onlinecite{DSFisher}], the collective variables representing moving (creeping) domain walls and/or interfaces between different phases is the profile of the interface.
\end{references}
\end{document}